\documentclass[11pt]{article}
\usepackage{moriond,epsfig}

\def\Journal#1#2#3#4{{#1} {\bf #2}, #3 (#4)}

\def\NPB{{\em Nucl. Phys.} B}
\def\PLB{{\em Phys. Lett.}  B}
\def\PRL{\em Phys. Rev. Lett.}
\def\PRD{{\em Phys. Rev.} D}
\def\PRDRC{{\em Phys. Rev.} D-RC}

\def\EPJ{{\em Eur. Phys. J.} C}
\def\JHP{\em JHEP}
\def\IJMP{{\em Int. J. Mod. Phys} A}

\def\be{\begin{equation}}
\def\ee{\end{equation}}
\def\bea{\begin{eqnarray}}
\def\eea{\end{eqnarray}}

\input babarsym

\def\er #1 #2 { $#1 \pm #2$ }

\def\vub      {\ensuremath {V_{ub}}}
\def\vcb      {\ensuremath {V_{cb}}}

\def\mpl #1 #2 #3 {Mod.~Phys.~Lett.~{\bf#1},\ #2 (#3)}
\def\npb  #1 #2 #3 {Nucl.~Phys.~B~{\bf#1},\ #2 (#3)}
\def\plb  #1 #2 #3 {Phys.~Lett.~B~{\bf#1},\ #2 (#3)}
\def\pr   #1 #2 #3 {Phys.~Rep.~{\bf#1},\ #2 (#3)}
\def\prd  #1 #2 #3 {Phys.~Rev.~D~{\bf#1},\ #2 (#3)}
\def\prl  #1 #2 #3 {Phys.~Rev.~Lett.~{\bf#1},\ #2 (#3)}
\def\RMP  #1 #2 #3 {Rev.~Mod.~Phys.~{\bf#1},\ #2 (#3)}
\def\zpc  #1 #2 #3 {Z.~Phys.~C~{\bf#1},\ #2 (#3)}
\def\nim  #1 #2 #3 {Nucl.~Instrum.~Methods~{\bf#1},\ #2 (#3)}
\def\nima  #1 #2 #3 {Nucl.~Instrum.~Methods~A.{\bf#1},\ #2 (#3)}
\def\epjc #1 #2 #3 {Euro.~Phys.~Jour~{\bf#1},\ #2 (#3)}
\def\rmp #1 #2 #3 {Rev.~Mod.~Phys~{\bf#1},\ #2 (#3)}
\def\npbps #1 #2 #3 {Nucl.~Phys.~B.~proc.~suppl~{\bf#1},\ #2 (#3)}
\def\progtp #1 #2 #3 {Prog.~Theo.~Phys~{\bf#1},\ #2 (#3)}

\newcommand {\bpigen} {\ensuremath{B \rightarrow \pi l \nu}\hbox{ }}

\newcommand {\bclnu}{\ensuremath{\b \rightarrow \c l \nu}\hbox{ }}

\newcommand {\Bxclnu}{\ensuremath{\Bb \rightarrow X_c \ell \bar{\nu}}}
\newcommand {\Bxulnu}{\ensuremath{\Bb \rightarrow X_u \ell \bar{\nu}}}

\newcommand{\beq}{\begin{equation}}
\newcommand{\beqa}{\begin{eqnarray}}
\newcommand{\beqn}{\begin{eqnarray}}
\newcommand{\eeq}{\end{equation}}
\newcommand{\eeqa}{\end{eqnarray}}
\newcommand{\eeqn}{\end{eqnarray}}
\makeatletter
\def\slash#1{{\mathpalette\c@ncel{#1}}} 
\makeatother

\begin{document}
\vspace*{4cm}
\title{SEMILEPTONIC B DECAYS AT THE B FACTORIES}

\author{ C. BOZZI}

\address{INFN Sezione di Ferrara, Polo Scientifico Tecnologico, Edificio C\\
Via Saragat, 1 - 44100 FERRARA, Italy\\
~  \\
~ \\
representing the Babar and Belle Collaborations}

\maketitle\abstracts{
Recent results on inclusive and exclusive semileptonic $B$ decays from $B$ 
Factories are presented. The status and perspectives of the determination 
of the CKM matrix elements $V_{ub}$ and $V_{cb}$ with semileptonic $B$ decays 
is discussed.}

\section{Introduction}

Semileptonic $B$ decays provide direct access to the CKM matrix elements \vub\ and \vcb, whose 
ratio gives a measurement of the side of the Unitarity Triangle opposite the angle $\beta$. 

The underlying theory of semileptonic $B$ decays is at an advanced stage. The weak 
currents factorize in leptonic and hadronic parts which do not interact between each other. 
Moreover, the $b$ quark mass is considerably larger than the scale $\Lambda_{QCD}$ of hadronic physics. 
Therefore, a systematic expansion in $\Lambda_{QCD}/m_b$ and $\alpha_s$ can be performed, 
contributions from perturbative and non-perturbative physics can be separated, and 
measurements of semileptonic $B$ decays enable precise determinations of \vcb\ and \vub.  

Semileptonic $B$ decays also probe the structure of $B$ mesons, in analogy with neutrino 
deep inelastic scattering. Inclusive decays are sensitive to quantities such as the 
mass and momentum distribution of the $b$ quark inside the $B$ meson, whereas exclusive decays 
measure form factors for specific final states. As a consequence, dominant theoretical uncertainties 
which enter in the determination of \vub\ and \vcb\ can be experimentally assessed and minimized.  

Due to large data sets, $B$ Factories are now performing precision measurements on high purity samples, 
such as events where one $B$ from the \FourS\ decay is fully or partially reconstructed, and the 
semileptonic decay of the other $B$ is studied. As a consequence, partial rates for \Bxulnu\ 
transitions can be measured in regions of the phase space, previously considered inaccessible, 
where theoretical uncertainties are reduced. 

A comprehensive review of \vcb\ and \vub\ measurements is beyond the scope of this paper. 
In the following, after discussing the general framework of each measurement technique, 
we will present and discuss only the most recent determinations of \vcb\ and \vub\ with inclusive 
(Sections \ref{sec:vcbincl} and \ref{sec:vubincl}) and exclusive (Sections \ref{sec:vcbexcl} 
and \ref{sec:vubexcl}) semileptonic $B$ decays. The current status and outlook are summarized 
in Section \ref{sec:concl}. 

\section{Inclusive Semileptonic Decays with Charm} \label{sec:vcbincl}

The rate for \Bxclnu\ decays is related to the free quark rate by Operator Product 
Expansion (OPE) techniques~\cite{ref:OPE}. The resulting {\em{Heavy Quarks Expansion}} is double 
in powers of $\alpha_s$ and $1/m_b$, and can be written schematically as~\cite{ref:OPEvcb}  
\beq \label{eq:vcbincl}
\Gamma_{sl} = \frac{G_F m_b^5}{192 \pi^3} |V_{cb}|^2 (1+A_{EW})A_{pert}(\alpha_s)
A_{non-pert}(1/m_b, 1/m_c, a_i)
\eeq
where $A_{EW}$ and $A_{pert}$ represent electroweak and QCD perturbative corrections 
respectively. The non-perturbative QCD part $A_{non-pert}$ is expanded in terms of the heavy quark 
masses ($m_b$, $m_c$), with $a_i$ as coefficients. The latter are matrix elements of operators, such 
as the kinetic energy and the chromomagnetic moment, which describe in principle universal properties 
of $B$ mesons. In practice, the $a_i$ parameters depend on the renormalization scale, on the chosen 
renormalization scheme, and their number is a function of the order of the $1/m_b$ 
expansion; four of them appear at order $m_b^{-3}$. 

The Heavy Quark Expansion also predicts the moments of observables in \Bxclnu\ decays such as the 
lepton energy $E_\ell$ and the invariant mass of the hadronic final state $m_X$, in regions of 
phase space, as a function of the $a_i$ parameters, $m_b$, $m_c$, and \vcb. Therefore, experimental 
determinations of these moments in different portions of the phase space allow a simultaneous 
measurement of the heavy quark parameters and masses, as well as \vcb. Such studies have been 
performed at the $B$ Factories~\cite{ref:CLEOmoments,ref:BABARmoments,ref:BELLEmoments}, 
CDF~\cite{ref:CDFmoments} and Delphi~\cite{ref:DELPHImoments}. 
The resulting moments of the energy spectrum (0th-3rd) and 
of the squared hadronic mass spectrum (0th-2nd) have comparable statistical and systematic 
uncertainties. 

Radiative \btosgam\ decays can also be used to determine the heavy quark parameters, since the 
energy spectrum of the photon, monoenergetic at the parton level, is smeared at the hadron 
level, by an amount which depends on the structure of the $B$ meson. The same techniques used 
in semileptonic decays can be applied in this case as well. Moments can be 
computed and compared to experimental determinations~\cite{ref:CLEObsg,ref:BELLEbsg,ref:BABARbsgexcl,ref:BABARbsgincl}. 
The main limitation of these measurements is due to background subtraction, which is dominant 
at low ($E_{\gamma}<1.9$GeV) energies. 

The experimental determinations of the \bclnu\ and \btosgam\ moments can be combined, 
by using the kinetic mass scheme, in a global fit to heavy quark 
parameters~\cite{ref:HQEfit}, which gives uncertainties of about 2\% on $|\vcb|$, 1\% on 
$m_b$ and 10\% on $\mu_\pi^2$, the matrix element of the kinetic energy operator: 
\bea \nonumber
& |\vcb| = (41.96 \pm 0.23 \pm 0.35 \pm 0.59) \cdot 10^{-3} & \\ 
m_b = (4.59\pm0.04) GeV & \mu_\pi^2 = (0.40 \pm 0.04) GeV^2 & \rho = -0.26 \label{vcbeq} 
\eea
Since $m_b$ and $\mu_\pi^2$ contribute the dominant systematic uncertainty in the inclusive 
determination of \vub, their precise measurement is crucial. 

\section{Inclusive Charmless Semileptonic Decays} \label{sec:vubincl}

In principle, $|\vub|$ is related to the rate of inclusive charmless semileptonic decays by 
an expression equivalent to Eq. \ref{eq:vcbincl}, which contains matrix elements of operators 
related to the ones entering in the \Bxclnu\ decay. 
If the full \Bxulnu\ decay rate were experimentally accessible, the resulting theory uncertainty 
would be of the order of 5\% ~\cite{ref:vub5pc}. In practice the accessible rate is much reduced and the theoretical 
uncertainty increases considerably, since the overwhelming background (a factor 50) from \Bxclnu\ 
decays must be suppressed by stringent kinematic requirements. These cuts are all based on the $u$ 
quark being much lighter than the $c$ quark. As a consequence, the distributions of $E_\ell$ and 
$q^2$, the squared invariant mass of the lepton pair, extends to higher values for signal, 
whereas the $m_X$ spectrum is concentrated at lower values. It is therefore possible to select 
regions of the phase space where the signal over background ratio is adequate. However, the 
resulting acceptances tend to be small (6\%, 20\%, up to 70\% for typical requirements on $E_\ell$, 
$q^2$ and $m_X$, respectively) and, if cuts are not carefully chosen, poorly known, since OPE 
breaks down and a shape function is needed to resum non-perturbative physics to all orders. 
This shape function depends on $m_b$ and heavy quark parameters. Therefore, it is possible to determine 
its basic features from other processes, like \Bxclnu\ and \btosgam\ decays~\footnote{Some care must 
be taken when relating the heavy quark parameters determined in different processes, since the 
theoretical calculations are performed in different normalization schemes, and the order of the 
heavy quark expansion is not necessarily the same.}. 
Indeed, most of the theoretical uncertainty in inclusive \vub\ determinations is due to 
our imperfect knowledge of the shape function, $m_b$ and the heavy quark parameters. Minimizing 
these uncertainties by maximizing the acceptance, {\it e.g.} by relaxing the cut on $E_\ell$, is 
possible only if background knowledge is good. Otherwise, one 
can choose regions, e.g. at low $m_X$ and high $q^2$, where shape function effects are expected 
to be small and OPE works well. 

Several theory calculations~\cite{ref:dfn,ref:bll,ref:blnp,ref:gardi} can be used to get 
acceptances in restricted regions of phase space. $|\vub|$ is determined from the measurement 
of $\Delta {{\cal{B}}(\Bxulnu)}$, the charmless semileptonic partial branching fraction in the phase space 
region $\Delta\Phi$ defined by kinematic cuts, and $\zeta(\Delta\Phi)$, the rate 
(in $|\vub|^2$ ps$^{-1}$) for the same phase space region predicted by theory: 
\beq
|V_{ub}| = \sqrt{\frac{\Delta {\cal{B}}(\Bxulnu)}{\tau_b \cdot \zeta(\Delta\Phi)}}
\eeq
where $\tau_b$ is the $B$ meson lifetime. Measuring partial branching fractions allows to 
use and compare several models, and to update previous determinations as theory uncertainties 
and calculations improve.  

\subsection{Measurements near the Endpoint of the Lepton Energy Spectrum}

\begin{figure}
\center{\psfig{figure=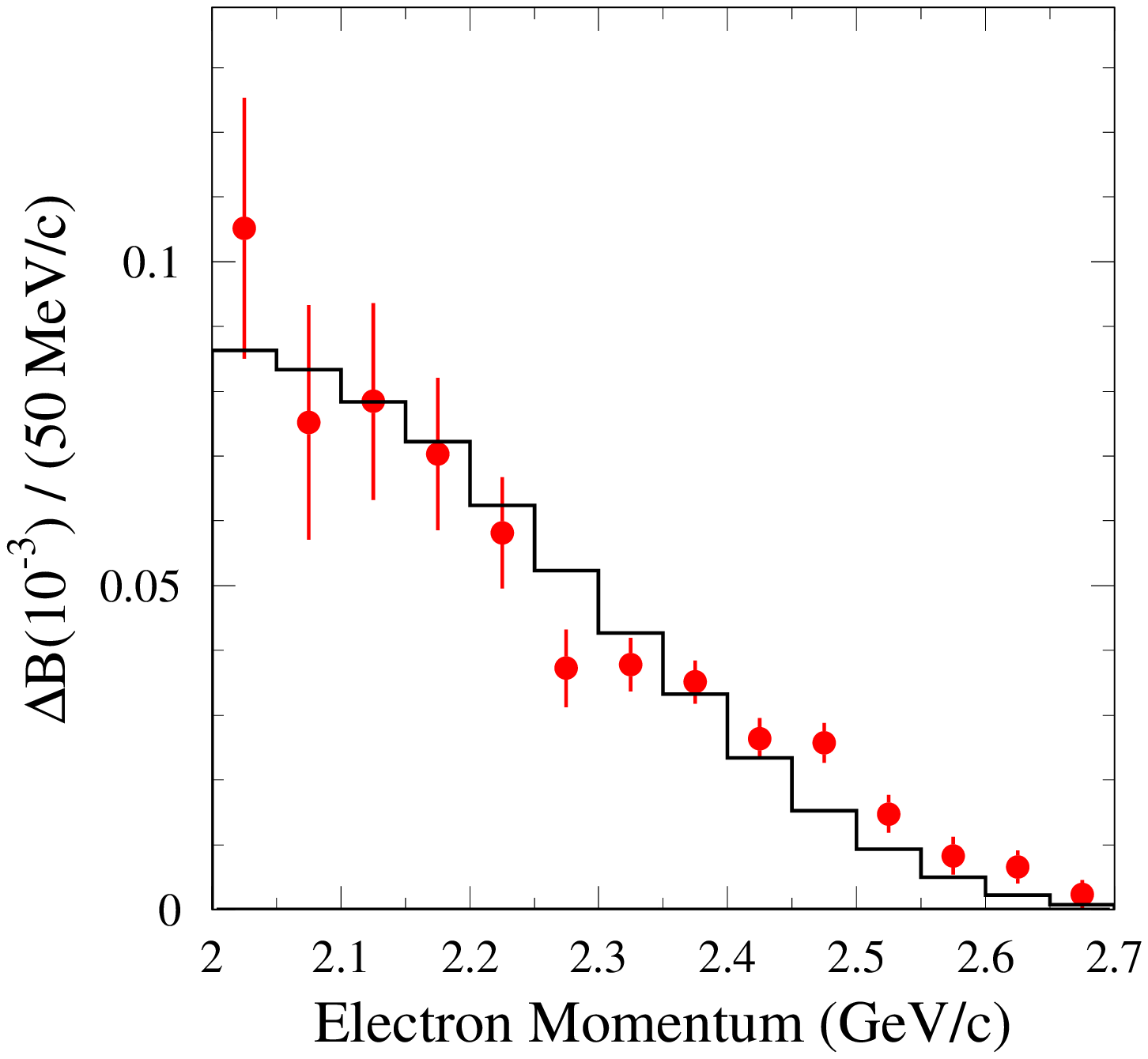,height=2.9in}
\psfig{figure=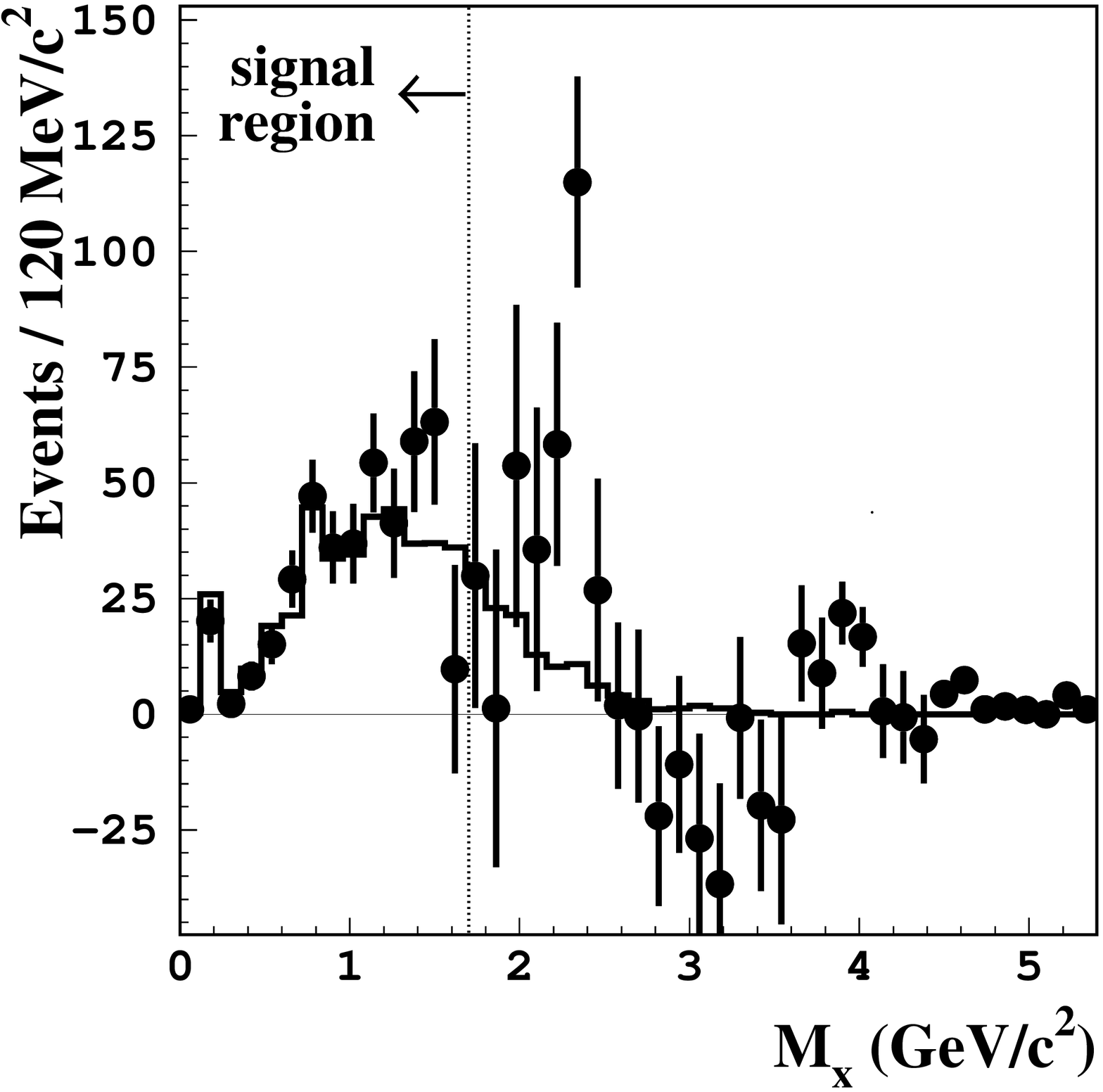,height=3in}}
\caption{Left: Electron energy spectrum near the endpoint, obtained by Babar on a sample 
of 88$\times 10^6$ \BB events. 
Data points, after background subtraction and correction for efficiency, 
bremsstrahlung  and final state radiation, are compared to the Monte carlo simulation 
(histogram). Right: Distribution of $m_X$,  
obtained by Belle on a sample of 253$\times 10^6$ \BB events. A 
$B$ meson is reconstructed in a fully hadronic final state and a semileptonic decay of the second $B$ 
is identified. Points represent data, after subtracting \Bxclnu\ and other backgrounds, the histogram is the fitted 
\Bxulnu\ contribution.
\label{fig:bbrendpoint}}
\end{figure}

Historically, the observation~ of events where the energy of the lepton 
exceeded the endpoint expected for \Bxclnu\ decays ($E_\ell > 2.3$ GeV) 
provided the first evidence for charmless semileptonic decays. 
As background knowledge improved, it has been possible~\cite{ref:cleo-endpoint,ref:babar-endpoint,ref:belle-endpoint} 
to relax the requirement on $E_\ell$ down to 2.0 GeV, or 
even 1.9 GeV (Belle), thereby increasing the acceptance and decreasing theory uncertainty. The 
typical signal-to-background ratio in these studies is about 
1:10. Figure \ref{fig:bbrendpoint}, left, shows the distribution of the electron momentum near 
the kinematic endpoint, after subtraction of backgrounds and corrections for efficiency and 
radiative effects, obtained by Babar on a sample of 88 million \BB events. The resulting 
determination of $|\vub|$, adjusted by HFAG~\cite{ref:hfag} (see Section \ref{vubres}), 
is shown in Table \ref{tab:inclvub} together with measurements from other experiments. The 
uncertainty on \vub\ from endpoint measurements is at the 10\% level, dominated by 
uncertainties on the shape function parameters, mostly a 40 MeV uncertainty on $m_b$. 

\subsection{Hadronic $B$ tags}

Other discriminating variables such as $m_X$ and $q^2$ can be reconstructed experimentally 
by determining  unambiguously which hadrons originate from the semileptonic decay of a $B$ meson. This 
difficult task for experiments running at the \FourS\ peak can be solved by explicitly 
reconstructing the decay of a $B$ meson in a fully hadronic final state, and studying the recoiling 
$B$, whose momentum and flavour are consequently known. 
If the recoiling $B$ decays semileptonically, the only missing particle in the event is the 
neutrino, which can be reconstructed by using missing mass arguments. The experimental resolution 
on the discriminating  variables can be increased by using a kinematic fit. 
This technique, pioneered by Babar~\cite{ref:bbrprl}, provides signal over background ratios of about 
1 or more, at the expense of a very small (${\cal{O}}(10^{-3})$) efficiency due to the full 
hadronic reconstruction. As datasets increase, inclusive measurements with hadronic $B$ tags are 
expected to give the most precise determinations of $|\vub|$. Figure \ref{fig:bbrendpoint}, 
right, shows the $m_X$ spectrum after background subtraction, resulting from an analysis 
by Belle~\cite{ref:belle-mx}. The resulting \vub\ measurement, adjusted by HFAG, is shown in 
Table \ref{tab:inclvub}, together with results from Babar. Results from the two experiments 
are comparable, with uncertainties at the 10\% level, dominated by theory. 

\begin{figure}
\center{\psfig{figure=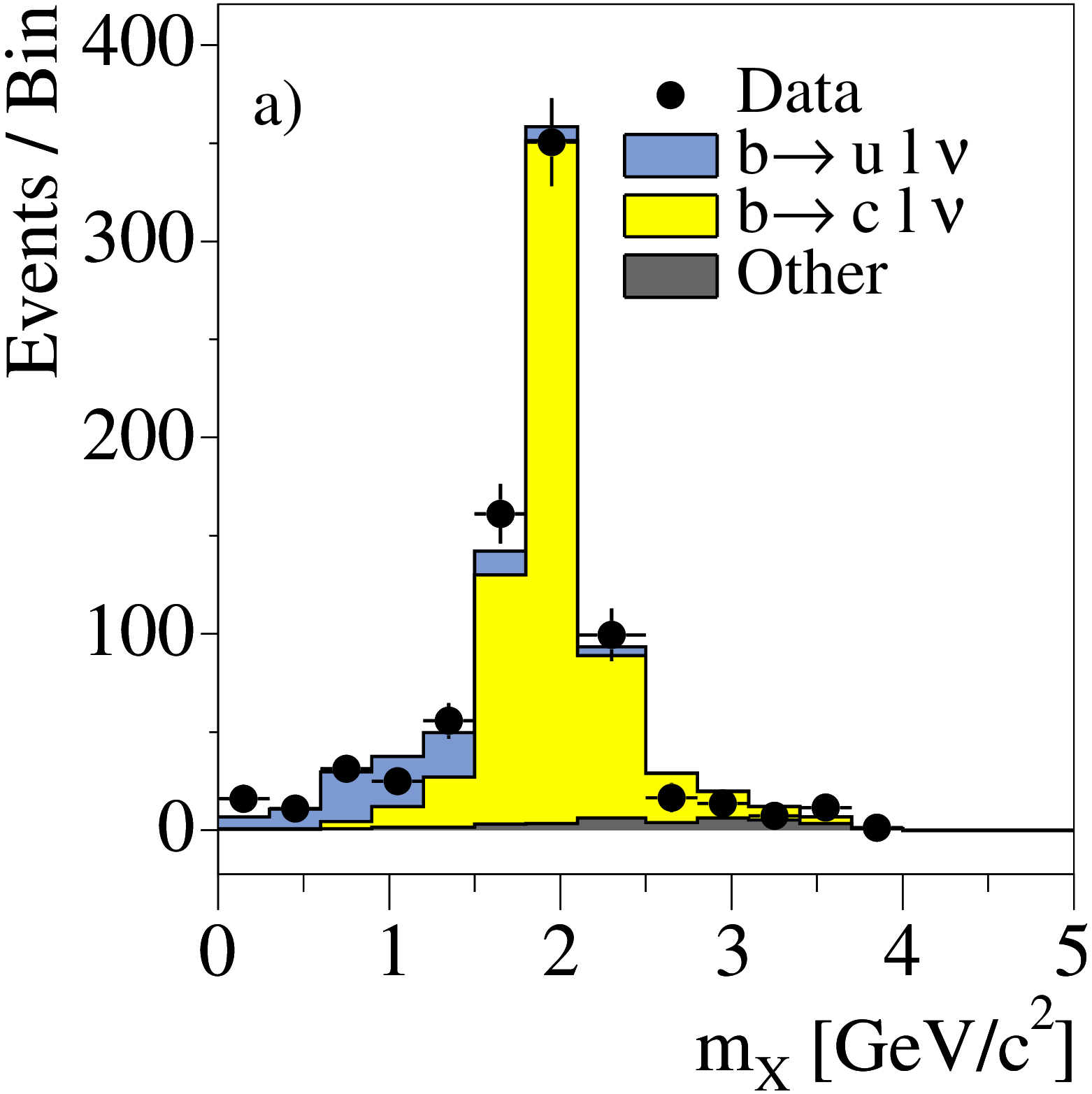,height=2.8in}
\psfig{figure=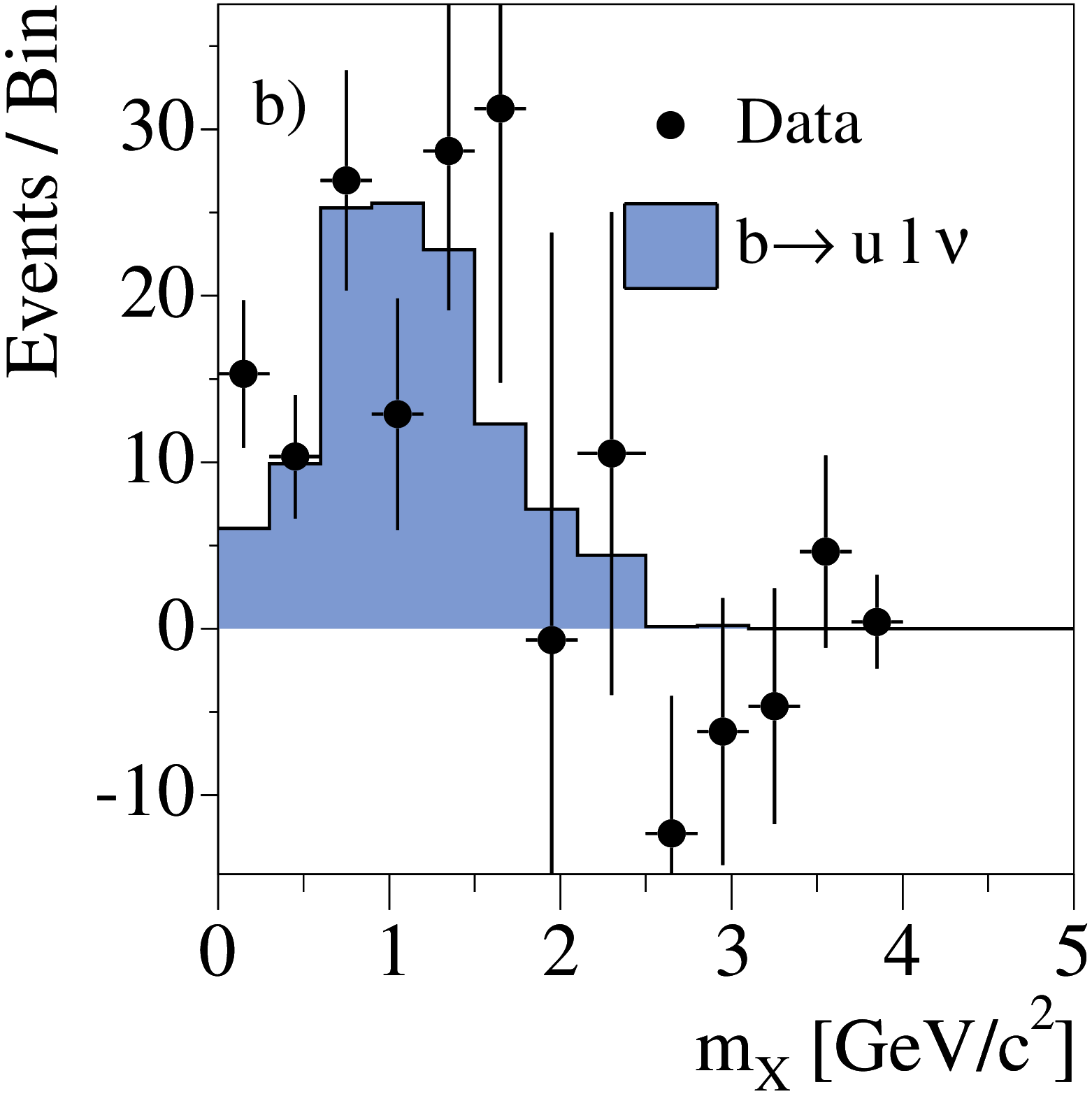,height=2.8in}}
\caption{Distributions of $m_X$ before (left) and after (right) subtraction of \Bxclnu\ and other  
backgrounds, obtained by Babar on a sample of 88$\times 10^{6}$ \BB events. Semileptonic decays 
are identified in the recoil of a fully reconstructed hadronic $B$ decay. Points are data, 
histograms represent signal and background contributions. The $m_X$ spectrum after background subtraction is used to 
compute partial branching fractions as a function of the $m_X$ cut, which are then combined with 
suitable integrations of the photon energy spectrum in \btosgam\ decays to obtain \vub. 
\label{fig:ed}}
\end{figure}

%
Shape function effects, which give the dominant contribution to the uncertainty in inclusive 
\vub\ determinations, can be reduced by using theoretical calculations which 
relate the rate $\Delta\Gamma(\Bxulnu)$ to the photon energy spectrum in 
\btosgam\ decays. For instance, one can schematically write~\cite{ref:neubnosf,ref:LLR}
\beq
\Delta\Gamma(\Bxulnu) = \frac{|V_{ub}|^2}{|V_{ts}|^2}\int{W(E_{\gamma})\frac{d\Gamma(\btosgam)}
{dE_{\gamma}}dE_{\gamma}}, 
\eeq
where the integration is performed in an appropriate phase space region and the weight 
function $W(E_{\gamma})$ is computed by theory with moderate uncertainty. 
The dependence on the shape function is therefore folded in the experimental measurement. 
A new Babar measurement~\cite{ref:ed}, which uses this approach and calculations by Low, Leibovich and 
Rothstein~\cite{ref:LLR}, has been released just before this Conference. It is 
based on an analysis of the recoil of fully reconstructed $B$ mesons on 88 million \BB events. 
Figure \ref{fig:ed} shows the spectrum of the hadronic invariant mass, before (left) and 
after (right) background subtraction.  
The partial rate $\Delta\Gamma(\Bxulnu)$ is determined by counting events below a cut on $m_X$, 
and \vub\ is extracted by using the photon energy spectrum in \btosgam\ decays as measured by 
Babar~\cite{ref:BABARbsgexcl}. Theory uncertainties increase as the $m_X$ cut is decreased, since 
a region dominated by non-perturbative effects is selected. Experimental errors 
increase at higher $m_X$ cuts due to background subtraction. The resulting optimal working point 
corresponds to $m_X<$ 1.67 GeV, which gives a measurement of $|\vub|$ (see Table \ref{tab:inclvub}) 
compatible with other determinations and 12\% uncertainty. As expected, the impact of shape function 
parameters is small in this approach. The theory error results from neglecting high 
order terms in the $1/m_b$ expansion. A \vub\ measurement over the full $m_X$ spectrum is also 
shown in Table \ref{tab:inclvub}; as expected, the theory error decreases since the full phase 
space is used, but the statistical error increases. 

\begin{table}[h]
\caption{Experimental measurements of partial branching
  fractions $\Delta\cal{B}$ for inclusive \Bxulnu\ decays and $|\vub|$, 
  adjusted by HFAG as explained in the text.   $f_u$ is the space phase acceptance. 
  The errors on \vub\ refer to experimental and theoretical uncertainties, respectively.  The
$s_\mathrm{h}^{\mathrm{max}}$ variable is described elsewhere$^{18}$. 
The $P_{+}$ variable is  defined as $P_{+} = E_X - |\vec{p}_X|$. }
\label{tab:inclvub}
\begin{center}
\begin{small}
\begin{tabular}{|l|l|c|c|c|}
\hline
& accepted region & $f_u$ 
& $\Delta{\cal{B}} [10^{-4}]$ & $\Vub [10^{-3}]$\\
\hline
$*$CLEO~\cite{ref:cleo-endpoint}
& $E_e>2.1\,\gev$ & 0.19
& $3.3\pm 0.2\pm 0.7$           & $4.05\pm 0.47\pm 0.36$ \\ 
$*$BABAR~\cite{ref:babar-endpoint}
& $E_e>2.0\,\gev$  & 0.26
& $5.3\pm 0.3\pm 0.5$           & $4.25\pm 0.30\pm 0.31$ \\ 
$*$BELLE~\cite{ref:belle-endpoint}
& $E_e>1.9\,\gev$  & 0.34
& $8.5\pm 0.4\pm 1.5$           & $4.85\pm 0.45\pm 0.31$ \\ 
$*$BABAR~\cite{ref:babar-elq2}
& $E_e>2.0\,\gev$, $s_\mathrm{h}^{\mathrm{max}}<3.5\,\mathrm{GeV^2}$ & 0.19
& $3.5\pm 0.3\pm 0.3$           & $4.06\pm 0.27\pm 0.36$ \\
$*$BABAR~\cite{ref:babar-q2mx}
& $m_X<1.7\,\gev/c^2, q^2>8\,\gev^2/c^2$ & 0.34
& $8.7\pm 0.9\pm 0.9$           & $4.79 \pm 0.35 \pm 0.33$ \\ 
$*$BELLE~\cite{ref:belle-mxq2Anneal}
& $m_X<1.7\,\gev/c^2, q^2>8\,\gev^2/c^2$ & 0.34
& $7.4\pm 0.9\pm 1.3$           & $4.41 \pm 0.46 \pm 0.30$ \\ 
BELLE~\cite{ref:belle-mx}
& $m_X<1.7\,\gev/c^2, q^2>8\,\gev^2/c^2$ & 0.34
& $8.4\pm 0.8\pm 1.0$           & $4.68 \pm 0.37 \pm 0.32$ \\
BELLE~\cite{ref:belle-mx}
& $P_+<0.66\,\gev$  & 0.57
& $11.0\pm 1.0\pm 1.6$          & $4.14 \pm 0.35 \pm 0.29 $ \\
$*$BELLE~\cite{ref:belle-mx}
& $m_X<1.7\,\gev/c^2$   & 0.66
& $12.4\pm 1.1\pm 1.2$          & $4.10 \pm 0.27 \pm 0.25$ \\ 
BABAR~\cite{ref:ed}
& $m_X<1.67\,\gev$ \& \btosgam & 
&  & $4.43 \pm 0.45 \pm 0.29 $ \\ 
BABAR~\cite{ref:ed}
& $m_X<2.5\,\gev$ \& \btosgam & 
&  & $4.34 \pm 0.76 \pm 0.10 $ \\ \hline \hline
Average of $*$ 
& $\chi^2=6.3/6$, CL=0.40 &
&                    & $4.39 \pm 0.19 \pm 0.27$ \\
\hline
\end{tabular}
\end{small}
\end{center}
\end{table}

\subsection{Inclusive \vub: Summary and Outlook} \label{vubres}

Table \ref{tab:inclvub} shows a summary of inclusive \vub\ measurements, together with the latest 
average from HFAG~\cite{ref:hfag}. All measurements have been adjusted by HFAG so that the same 
theory framework~\cite{ref:blnp} and shape function parameters and uncertainties 
(Eq. \ref{vcbeq}) are used. The total uncertainty on \vub\ from 
inclusive measurements is 7.4\%, dominated by theory. The uncertainty from 
limited knowledge of the shape function is about 4\%. Experiments can help by determining 
shape function parameters with better accuracy, but it will be hard to go down  
30 MeV on $m_b$. Other theoretical uncertainties, due to neglecting higher order 
terms and weak annihilation effects, contribute 5\% to the uncertainty on \vub. 
While the latter can be studied experimentally, the former will be difficult to improve. 
Other theory approaches, such as the Dressed Gluon 
Exponentiation by Andersen and Gardi~\cite{ref:gardi}, are also promising and worth to 
investigate. 

\section{Exclusive SL Decays with Charm} \label{sec:vcbexcl}

\begin{figure}[b]
\center{\psfig{figure=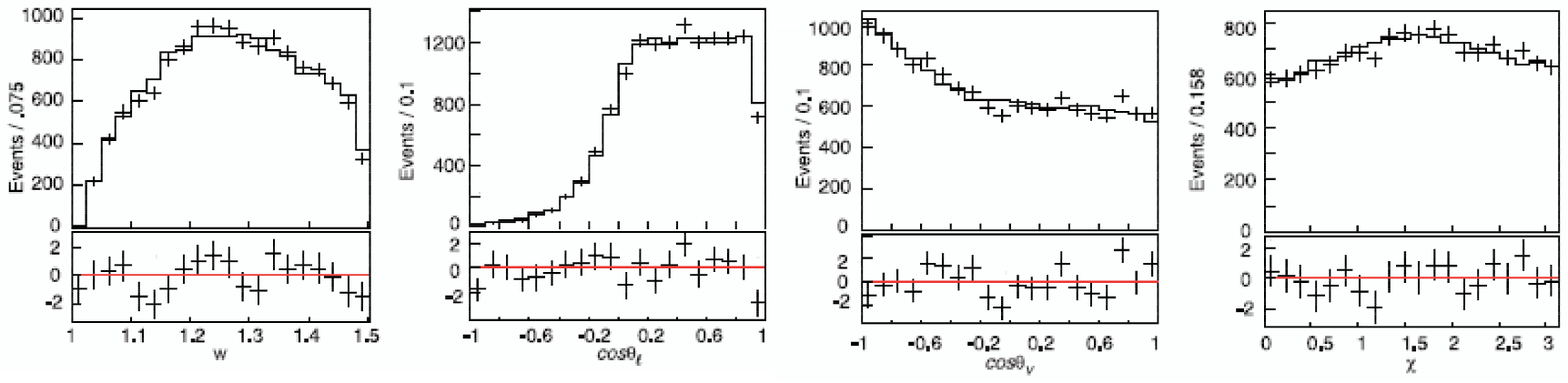,width=\textwidth}}
\caption{Background-subtracted data (points) overlaid on Monte Carlo (histograms) 
for the distributions of the four kinematic variables relevant in the \Btodstlnu\ decay, 
as measured in the Babar form factor analysis on a sample of 86$\times 10^6$ \BB events. 
Simulated events have been reweighted according to the best fit of the form factors. The bottom panel of each figure shows 
the pull (difference over error) plot. The line at zero is shown for comparison purposes only. 
\label{fig:mandeep}}
\end{figure}

The technique of determining \vcb\ by using \Btodstlnu\ decays is well established. 
The differential distribution can be written in terms of $w$, the $D^*$ boost in the $B$ rest 
frame, as 
\beq
\frac{d\Gamma(\Btodstlnu)}{dw} = \frac{G_F^2 |\vcb|^2}{48 \pi^3} ({\cal{F}}(w))^2 {\cal{G}}(w)
\eeq
where ${\cal{G}}(w)$ is a phase space factor and ${\cal{F}}(w)$ is a form factor which would be 1 
at $w=1$ in the heavy quark limit. Lattice QCD can be used to compute effects due to finite quark 
masses, leading to~\cite{ref:hashimoto} ${\cal{F}}(1)=0.919 ^{+0.030}_{-0.035}$. The shape of 
${\cal{F}}(w)$ cannot be predicted by theory, and is parameterized in terms of a slope $\rho^2$ and 
form factor ratios $R_1$ and $R_2$, (nearly) independent of $w$. The helicity amplitudes entering 
in the \Btodstlnu\ decay are also function of the above parameters. These amplitudes, and 
therefore $\rho^2$ $R_1$ and $R_2$, can be determined by fitting the four-fold 
differential rate of \Btodstlnu\ decays in terms of $w$ and three angles which describe 
the decay kinematics. Figure \ref{fig:mandeep} shows the result of the fit to the angular 
and $w$ distributions obtained in a recent Babar measurement~\cite{ref:mandeep}, where form 
factors are parameterized by using a prescription due to Caprini, Lellouch and 
Neubert~\cite{ref:CLN}. 
The uncertainties on the resulting measurements  
\bea \nonumber
R_1 = 1.396 \pm 0.060 \pm 0.044, &
R_2 = 0.885 \pm 0.040 \pm 0.026, &
\rho^2 = 1.145 \pm 0.059 \pm 0.046, 
\eea
are a factor 5 better than in previous determinations ~\cite{ref:CLEOFF}. Consequently, 
the systematic uncertainty, due to form factor 
ratios, in the Babar exclusive \vcb\ determination~\cite{ref:BBRexclvcb} 
decreases approximately by the same amount. It is also interesting to note that the 
re-interpretation of the Babar exclusive \vcb\ measurement gives 
\beq
|\vcb| = (37.6 \pm 0.3_{stat} \pm 1.3_{syst} \pm 1.4_{theory}) \times 10^{-3}, 
\eeq
which is about 2 standard deviations away from the published result~\cite{ref:BBRexclvcb}. 
Since \Btodstlnu\ is a dominant background for charmless semileptonic decays, a reduction 
of the systematic uncertainty due to the better knowledge of the \Btodstlnu\ form factor ratios 
is also observed in the endpoint measurement~\cite{ref:babar-endpoint} of \vub. 

\section{Exclusive Charmless SL Decays} \label{sec:vubexcl}

The differential rates for exclusive charmless semileptonic decays in terms of $q^2$ 
is proportional to $|\vub|^2$ times a form factor which is final state dependent. The absolute 
values of these form factors are predicted by using several theoretical frameworks (light-cone 
sum rules, lattice calculations, quark models); their dependence on $q^2$ can be checked 
experimentally, thereby allowing to discriminate different theoretical models. In brief, 
experiments search for semileptonic decays with a light meson ($\pi, \rho, \eta, \eta ', \omega$) 
in events where the other $B$ is tagged via hadronic or semileptonic decays, or even in untagged 
events. The latter gives better efficiencies, but also higher backgrounds. 
No new results were released immediately before this Conference. A summary of determinations of 
exclusive charmless semileptonic branching fractions and \vub\ is given elsewhere~\cite{ref:hfag}. 
The uncertainty on the average value of $|\vub|$, about 14\%, is dominated by the normalization 
of the form factors, which contributes about 10\%. The determinations of \vub\ with inclusive and 
exclusive decays are in agreement at the present level of accuracy. 

Heavy quark symmetry relates the form factors of \bpigen and $D\rightarrow \pi \ell \nu$ decays. 
A precise measurement of the latter represents a stringent test which can be used to calibrate 
theoretical calculations and increase the precision of \vub\ determinations from \bpigen 
decays. A first step towards this goal is measuring the $q^2$ dependence of the hadronic form factor 
in $D\rightarrow K \ell \nu$  decays with great accuracy. A preliminary result obtained by Babar 
is shown for the first time at this Conference. A sample of 2$\times 10^{5}$ decays has been analysed, and the 
$q^2$ distribution, unfolded of detector effects, has been obtained and fit to two different 
ans\"atze (the pole and modified pole mass) for the form factor shape. Measurements of the mass and 
scale entering in these parameterizations give uncertainties which are at least a factor 2 better 
than previous determinations. 

\section{Conclusion} \label{sec:concl}
The study of semileptonic $B$ decays is a very active area for both theory and experiment. 
Substantial progress has been obtained by applying HQE fits to inclusive \Bxclnu\ decays, 
resulting in precise measurements of $|\vcb|$ (2\%), $m_b$ (1\%) and heavy quark parameters 
relevant also to charmless decays. Increased precision (7\%) has been obtained in inclusive 
\vub\ determinations, by improving the existing techniques, using a more comprehensive 
theoretical treatment, and improving the determination of the $b$ quark mass. 

New precision measurement of form factors in exclusive \Btodstlnu\ decays allows to reduce 
systematic uncertainties in the determination of \vcb\ with exclusive decays and \vub\ with 
inclusive decays.  
Studies of exclusive charmless decays will improve as datasets increase. However, reducing the 
theoretical uncertainties to a level comparable with the statistical error and the inclusive 
determinations is challenging. In this respect, measurements of related processes such as 
semileptonic decays of charm mesons will increase confidence in theoretical calculations and 
uncertainties. 
%
%
%
%
%

\end{document}